%% file: main.tex
\documentclass[english,oneside,twocolumn]{article}
\usepackage[table]{xcolor}

\usepackage{listings}
\usepackage{bytefield}

\usepackage{amsfonts}

\usepackage[utf8]{inputenc}
\usepackage[english]{babel} 
\makeatletter
\def\bbl@iflanguage#1{\@ifundefined{l@#1}{\@gobble}\@firstofone}
\makeatother
\usepackage[big]{dgruyter_NEW}

\begin{document}

  \author[1]{Alex Catarineu}
  \author[2]{Philipp Claßen}
  \author[3]{Konark Modi}
  \author[4]{Josep M. Pujol}
  \affil[1]{Cliqz, E-mail: alex@cliqz.com}
  \affil[2]{Cliqz, E-mail: philipp@cliqz.com}
  \affil[3]{Cliqz, E-mail: konark@cliqz.com}
  \affil[4]{Cliqz, E-mail: josep@cliqz.com}

  \title{\huge Preventing Attacks on Anonymous Data Collection}

  \runningtitle{Preventing Attacks on Anonymous Data Collection}


\abstract{
\input{abstract.tex}
}





\maketitle




\input{intro_0.tex}

\input{preliminaries.tex}

\input{design.tex}

\input{protocol.tex}

\input{implementation.tex}

%

\input{evaluation.tex}

%

\input{conclusion.tex}

\clearpage

\def\bibfont{\normalsize}

\bibliographystyle{abbrv}

\bibliography{references}  
%

\clearpage

\appendix

\input{annex_1.tex}

\end{document}

%% file: abstract.tex
Anonymous data collection systems allow users to contribute the data necessary to build services and applications while preserving their privacy. Anonymity, however, can be abused by malicious agents aiming to subvert or to sabotage the data collection, for instance by injecting fabricated data.

In this paper we propose an efficient mechanism to rate-limit an attacker without compromising the privacy and anonymity of the users contributing data. The proposed system builds on top of Direct Anonymous Attestation, a proven cryptographic primitive. We describe how a set of rate-limiting rules can be formalized to define a normative space in which messages sent by an attacker can be linked, and consequently, dropped. We present all components needed to build and deploy such protection on existing data collection systems with little overhead.

Empirical evaluation yields performance up to 125 and 140 messages per second for senders and the collector respectively on nominal hardware. Latency of communication is bound to 4 seconds in the $95^{th}$ percentile when using Tor as network layer.








%% file: intro_0.tex
\section{Introduction}

It is common for a service to collect data from its users. Be it directly, for instace by gathering book ratings to offer recommendations on what to read next. Or indirectly, by tracking the user across to serve \textit{``tailored''} advertisement. Data collection is pervasive. Unfortunately, the standard methodology of collecting data poses a serious threat to the users' privacy.

Organizations that collect data, let us call them \textit{collectors}, are in general trustworthy, law abiding and with comprehensive data management and privacy policies. Despite good faith, collectors tend to accumulate all user's activity under a large profile, effectively linking all user's records together using an anchor. This methodology is problematic in regard to privacy for multiple reasons: hacks leading to a data breach~\cite{data:breaches}; disgruntled or unethical employees using data for their own benefit~\cite{leak:insider}; companies going bankrupt and selling the data as assets~\cite{leak:bankrupt}; government-issued subpoenas and backdoors~\cite{legal:theguardian, legal:verizon} are just some examples of the risks to which users are exposed when large profiles exist.

To collect a single data record is not so much of a problem with respect to privacy, but once records can be linked to a user, serious concerns arise. Let us illustrate it with an simple example. Three different GPS locations such as a home address, a work address and an kindergarten address. These three location records are innocuous in isolation, but if one knows that all three belong to the same user it is a totally different story. The user might be de-anonymized~\cite{narayanan2008robust, aol:privacy, demontjoye2013unique} and consequently his full location history exposed. Privacy is lost not by sending location records, but by the ability of the collector to link them altogether thanks to an anchor: be it a user-id, session-id, fingerprinting, etc. 

Why do organizations gather user's data in a linkable form? The answer to this question is out of the scope of the paper, but experience tells us it is mostly about convenience. Linkable data can repuposed to serve a wide-range of different services and applications, so it is natural to be the preferred format, despite privacy side-effects. 
Fortunately, this mindset is changing. There are some examples~\cite{humanweb, brave:bat} of data collection systems that prevent record linkability, thus, achieving true anonymity for users contributing data. \textbf{In an anonymous data collection setup the collector has no means to determine whether two records belong to the same person.}



\clearpage

\subsection{Problem Statement}

Anonymous data collection come with the following problem: \textbf{if the user is fully anonymous, how can the system prevent an attacker from polluting the data collection?}

This problem is not exclusive to the anonymous data collection methodology, but it is exacerbated when anonymity is a must.

Conventional collectors also suffer from data pollution attacks, e.g. trying to alter ranking on search engines ~\cite{gyongyi}, fraudulent clicking on ads~\cite{daswani, dave}, etc.
However, in the case of conventional collectors, record-linkage is not forbidden, as a matter of fact it is encouraged. User identifiers, pseudonymous or not, can be leveraged to defend against and attacker via Authentications schemes, API keys or via statistical or machine-learning outlier detection~\cite{abbott, kitts} techniques. 

Unfortunately, when users are truly anonymous one cannot rely on defenses that are predicated on the ability to link records to their origin. Not being able to detect attackers who take advantage of anonymity is a big hurdle for the deployment of anonymous data collection at scale.

\subsection{Related Work}

Deployed anonymous data collection systems are scarce, we hope to witness more of them as organizations re-evaluate their stance on the trade-off between convenience and privacy.

We should emphasize the work by Anonize~\cite{anonize15}, which focuses on providing anonymity to users filling surveys. This system allows the creation of ad-hoc groups of users (cohorts) who can submit the same survey (data record) only once. To build on top of Anonize we would require more flexibility for limiting the number of records a user can send. It is unclear whether this would be achievable without modifying their protocol in a non-trivial way. The ad-hoc group selection is also not a requirement in our case, which allows us to save complexity.


Camenish \textit{et al.}~\cite{Camenisch:2006:WCE:1180405.1180431} presented an anonymous credential system that lets users authenticate at most $n$ times. This is a similar use-case as ours. We believe this system could be adapted to fulfill our rate-limiting needs. However, there has been many improvements since the work was first proposed, in the current state-of-the-art more efficient schemes exist which provide similar capabilities (e.g. relying on pairings instead of RSA). One instance is Direct Anonymous Attestation, which is the one we have finally chosen for our construction, the main reasons being that it has been extensively reviewed, has been standardized (or is in process) and has efficient implementations~\cite{cryptoeprint:2015:1246, lindemann_camenisch_drijvers_edgington_lehmann_urian, 10.1007/978-3-642-13869-0_12}.



Note that our work does not belong to the context of dataset anonymization, e.g. $k$-anonymity~\cite{kanonymity}, differential privacy~\cite{diffprivacy}, $t$-closeness~\cite{kanonymity:tcloseness}, homomorphic encryption~\cite{gentry2009fully}, etc. The scope is not on anonymizing data collection, but rather enabling anonymous data collection.



\subsection{Contributions} \label{section:contribution}

In this paper we present an expressive and efficient mechanism to \textbf{limit} the attackers capability to inject fabricated data into an anonymous data collection system, \textbf{without compromising the users' anonymity and privacy}.

There are some prior assumptions,

\begin{itemize}

\item We assume that the data sent by users is already anonymous. Data records do no contain any personal identifiable information (PII) or any other element that would allow the collector to link records coming from the same user.

\item We assume that a safe anonymous communication layer between user and collector already exists, for example the Tor network~\cite{tor:tor}. 

\item We assume that Sybil attacks~\cite{sybil} are not cost-effective. The system presented in this paper protects against attackers controlling one (or very few) users with valid credentials. Protection will degrade if an attacker is able to create a large number of sybils. We do not focus on preventing sybil attacks, but we do take some steps to mitigate them in Sect.~\ref{keyrotation}.

\end{itemize}

The system that presented in this paper builds on top of Direct Anonymous Attestation (DAA), a well-reviewed cryptographic primitive. We provide a flexible way of defining a normative space (via rate-limiting rules) that is enforced by the controlled linkability of DAA. The end result is that messages from attackers, who do not abide by the defined norms, will be linkable, and consequently, detected and dropped.








In Sect.~\ref{design} all aforementioned terms are properly defined. In Sect.~\ref{sec:examples} we provide some descriptive high-level examples for illustration purposes. In Sect.~\ref{sec:protocol} we describe the protocol on top of DAA. Evaluation of the end-to-end system performance on real users, using Tor as anonymization network, is presented in Sect.~\ref{evaluation}.

%% file: preliminaries.tex
\section{Preliminaries} \label{preliminaries}

In this section we recall some cryptographic primitives and notions.

\subsection*{Direct Anonymous Attestation}

Direct Anonymous Attestation (DAA) is a cryptographic primitive which enables anonymous remote authentication of a trusted computer. In some of its variants, it allows adding a string called \textit{basename} to achieve controlled linkability. Under this definition, two messages signed by the same user will be unlinkable \textbf{if and only if} their \textit{basename} is different.

We want to emphasize the importance of the \textit{basename} concept: this is the key idea that will allow us to achieve protection via rate-limiting in an anonymous way. How to use signatures and basenames to prevent data collection attacks will be detailed in section \ref{design}.

Let us introduce the Direct Anonymous Attestation (DAA) algorithms as defined in~\cite{10.1007/978-3-642-13869-0_12} to offer an overview of the operations relevant for the purposes of our system:

$\textbf{Setup}$: A randomized algorithm that produces a pair ($gpk$, $isk$), where $gpk$ is the group public key and $isk$ the issuer secret key.

$\textbf{Join}$: An interactive protocol run between a signer and an issuer $I$. At the end of the protocol, signer obtains a secret key $usk$ and membership credential $ucred$ issued by $I$.

$\textbf{Sign}$: On input of $gpk$, $usk$, $ucred$, a basename $bsn$, and a message $m$, the signer uses this algorithm to produce a signature $\sigma$ on $m$ under $(usk, ucred)$. The basename $bsn$ is a string used to control the linkability.

$\textbf{Verify}$: On input of $gpk$, $bsn$, $m$, a signature $\sigma$ on $m$, a verifier uses this algorithm to determine whether $\sigma$ is valid.

$\textbf{Link}$: On input of two signatures $\sigma_1$ and $\sigma_2$, a verifier uses this algorithm to determine whether the signatures are linked, unlinked or invalid.

For efficiency purposes, we require an additional \textbf{ExtractTag} algorithm, such that for two \textbf{valid} signatures $\sigma_1$ and $\sigma_2$:

\begin{equation*}
\begin{split}
    \boldsymbol{Link}(\sigma_1, \sigma_2) = linked \iff \\ \boldsymbol{ExtractTag}(\sigma_1) = \boldsymbol{ExtractTag}(\sigma_2) \\
\end{split}
\end{equation*}


It is important to remark that we do not consider a \textit{Trusted Platform Module} (TPM) that can be used to prove authorization to the issuer. Instead, we will assume that this authentication can be achieved via some long-lived keypair, which would serve as the user identity. 

Besides, we have to note that contrary to a DAA scheme, we do not specify that the basename $bsn$ has to be either a special symbol or the name string of an issuer. Instead we will use the basename to define the rate limits that construct the normative space, as we will show in the following sections.


\subsection*{Format-preserving encryption}

Format-preserving encryption~\cite{bellare2009format} allows encrypting in such a way that the ciphertext is in the same format as the plaintext. This means, for example, that if the plaintext is a always an integer between 0 and N - 1, then the ciphertext will also be.

For our purposes, we will define $\textit{FPE}_N(key, value) \rightarrow \left \{ 0, \ldots, N - 1 \right \}, key \in \left \{0, 1 \right \}^\tau, 0 \leq value < N$ as a function that encrypts an integer $value$ preserving its format. We will also assume that that the computation time for a single value $FPE_N(key, value)$ is constant.

Under this definition, we can use $FPE_N$ and a fixed random $key$ to generate a pseudo-random permutation of $\left\{0, \ldots, N - 1\right\}$ by encrypting all possible values: $\big( \textit{FPE}_N(key, 0), \textit{FPE}_N(key, 1), \ldots, \textit{FPE}_N(key, N - 1) \big)$.

%% file: design.tex
\section{Design} \label{design}

In this section we present the design of our system. By employing an anonymous credential system based on DAA, and its controlled linkability features, we will be able to detect and filter out messages from users who are not normative, i.e. that have exceeded the assigned quota.

In order to better understand the problem, let us consider three actors: a \textit{Client}, which executes in a user machine (e.g. a browser), a \textit{Collector}, some service that needs to receive some data from clients, and the  \textit{Verifier}, responsible to decide, for each input message, whether it is \textit{valid} (and forward it to the collector) or not (drop it). The \textit{Verifier} and \textit{Collector} can be the same entity.



For a message to be framed as \textit{valid} by the collector we want to impose several conditions:

\begin{enumerate}
    \item Message is \textit{well-formed}: it must belong to the set of valid messages as defined by the concrete system. In practice, this is equivalent to the server-side validation that most services do on user input data.
    \item User sending the message is authorized: has possession of some credentials that we have explicitly allowed.
    \item The message does not exceed some defined user quota.
\end{enumerate}

These conditions are straightforward to enforce if privacy preservation is not a requirement. For example, by forcing the user to attach his identity (public key) and a signature to every message the collector would be able to keep track of the full history of messages for every user. This way, doing any arbitrarily complex per-user rate-limiting would be a trivial task.

But we want achieve these goals in a privacy preserving way. We will show it is possible to do so by using a Direct Anonymous Attestation (DAA) cryptographic primitive.

Just by directly employing DAA we can already fulfill one of the conditions we wanted to enforce: ensuring that every message was sent by an authorized user, anonymously.

However, if an authorized user decides to subvert the system in some way and start sending malicious messages, it would still be impossible to detect, precisely because of the anonymity of the authentication system. Therefore, we want to be able to further limit the capabilities of users even if they are authorized.

Fortunately, DAA also allows controlled linkability via a basename string ($bsn$) that can be attached to a signature. More concretely, two messages signed with the same user credentials will be unlinkable \textbf{if and only if} their basenames are different. This feature can be instrumented to ensure that authorized users abide by some defined rules when sending messages to the collector.

We will present several ideas on how the structure of these basenames can be defined to achieve common rate-limiting patterns. We will proceed in an incremental fashion, starting with the simplest structure and moving step by step to achieve more expressive rate-limiting rules. Our final general construction will serve as a formalization for the concept of \textbf{rate-limiting ruleset}. At a high level view, each of these final rules will comprise three dimensions:

\begin{enumerate}
    \item A component that depends on the message content, called the \textbf{digest}.
    \item A \textbf{monitoring period} $K$, meaning that the rule will be \textit{reset} every $K$ units of time.
    \item A \textbf{multiplier} $N$, meaning that for any other fixed two dimensions the user will be able to send $N$ messages.
\end{enumerate}

\subsection{Rate-limiting rules construction}

\subsubsection{N-times anonymous authentication} \label{ntimesauth}

Consider the following structure for a basename:

\[ bsn \leftarrow \left< nonce \right>, 0 \leq nonce < N, N \in \mathbb{N} \]

With these constraints in place there are only $N$ possible distinct basenames that a user can construct. If a user sends $N + 1$ messages then two of them must use the same basename and therefore will be linkable by the collector. This allows us to easily filter out these extra non-allowed messages, effectively enforcing a simple rate-limiting rule that caps the number of messages an authorized user will be able to send.

While this can be useful in some cases, we would like to achieve more expressiveness in our rate-limiting rules.

\subsubsection{Limiting by time period} \label{timeperiodlimit}

A reasonable extension to the previous scheme is to include a timestamp with specific resolution to the basename:

\[ \left< \left\lfloor \frac{time}{K} \right\rfloor, nonce \right>, 0 \leq nonce < N \]

Here $time$ would be an integer indicating when the user sent the message and $K$ would be the monitoring period, in the same units as the $time$ (chosen depending on the application, e.g. hours). This limits the amount of messages a user can send to $N$ every $K$ units of time, which is slightly more general than the case seen in \ref{ntimesauth}.

\subsubsection{Message-specific rate-limiting} \label{messagespecific}

In order to achieve even more expressive power, we want to make the rate-limiting logic message specific. Besides, we would like to be able to enforce more than one rule at the same time. For example, we might want to have a rule to limit to $N_{day}$ the total number of messages a user can send per day and at the same time another rule that limits sending a specific class of message to $N_{class}$ every hour.

We can use the same DAA primitive to achieve this by attaching more than one signature per message, each one with a different basename, corresponding to each rate-limiting rule. Taking this into consideration, the final construction for the basenames would be as follows:

\[ bsn_i \leftarrow \left< digest_i(m), \left\lfloor \frac{time}{\textit{FK}_i(m)} \right\rfloor, nonce_i \right> \]
\[0 \leq nonce_i < \textit{FN}_i(m) \]

Note that by replacing the $N$ and $K$ constants by functions $\textit{FN}$ and $\textit{\textit{FK}}$ we allow the limits to depend arbitrarily on the message (based on content, type, etc.). Additionally, the $digest$ function allows the possibility for parts of the message content to be part of the rate-limiting rules. 

By allowing multiple basenames we can effectively enforce several rules at the same time if required.  Furthermore, we do not lose generality: we can still create rules with the semantics of sections \ref{ntimesauth} and \ref{timeperiodlimit}, by making the $digest_i$ return a constant and making $\textit{FK}_i$ return a very high number, so that the time component of the basename never changes.

We can finally model our \textbf{rate-limiting rules} as a list of triples of functions: $(digest_1, \textit{FN}_1, \textit{FK}_1), \ldots, (digest_n, \textit{FN}_n, \textit{FK}_n)$.

We refer to section \ref{sec:examples} for practical examples that make use of the possibilities that these general rate-limiting rules offer.

\subsection{Choosing the nonce} \label{nonce}

Let us define a \textit{pre-basename} as a basename with all fixed components except the nonce. How to choose a fresh nonce will depend on two factors: how many times that pre-basename has already been used and the specific rule limit $\textit{FN}_i(m)$.

A user should never send two messages with the same basename, otherwise they will be linked and blocked. Therefore, some state will need to be maintained so that a client can efficiently choose a random nonce from the unused ones, or abort if it has exceeded some quota. Note that it is important that for a given pre-basename, the sequence of nonces that the user selects is indeed random, to minimize the amount of information that a collector might have to attempt deanonymization.

By using \textit{Format-preserving encryption} we can achieve this in constant time (per message) and space. We only need to store a random $key$ and the number $n$ of messages that have already been sent, for every pre-basename. Then, in order to pick a fresh nonce it suffices to perform FPE encryption with the corresponding $key$ and $n$, and then increment $n$ for the corresponding pre-basename in the user storage, so that we can efficiently mark the nonce as used.

\subsection{User identities and key rotation} \label{keyrotation}

If we recall the \textbf{Join} protocol in DAA, the Issuer can communicate with a TPM to verify that the signer platform is entitled to receive anonymous credentials (join the group). 

We do not consider a TPM and therefore the problem of verifying user identities in the \textbf{join} operation becomes slightly more complex. We assume the existence of some long-lived public key (e.g. RSA, ECDSA, EdDSA,...) that serves as user identity, proven via signature. Depending on the use case, it might be possible to additionally require stronger means of identification, such as e-mail, mobile number, social security numbers, or even proof of work~\cite{back2002hashcash}, etc. The more difficult to generate or counterfeit, the better, since it will increase the cost of creating artificial users. It seems reasonable to assume that it will always be feasible for attackers to create multiple identities to gain an advantage for injecting malicious data (Sybil attack). We do not claim to avoid these attacks, but we take some reasonable steps in order to mitigate them.

First, we perform \textbf{periodic Issuer key rotation}, forcing users to obtain new credentials every so often. Then, we try to make it difficult for an attacker with many identities to renew credentials for all of them. In other words: we try to make it difficult to \textbf{Join} multiple times for an attacker. Note that the latter mechanism cannot be effective without the first one, otherwise an attacker would join once and then be able to send messages forever without additional effort.



Note that the \textbf{Join} operation does not need to be network-anonymized, the Issuer needs to see the long lived user public key, which is a stronger identifier than the IP address. Taking this into consideration we can leverage on the literature for anomaly and outlier detection. The difference is that we would not apply the detection methods on the data itself but on the \textbf{Join} operation. So anonymity of data is not compromised.



A side-effect of doing periodical key rotation is that both the user and the verifier can empty their stored pre-basenames or tags. This is so because it is ensured that different user private keys and user credentials will always produce different (unlinkable) signatures, even for identical basenames. This fact also has an important practical implication that must be taken into consideration: the rate-limiting ruleset \textbf{can only be enforced} for as long as the group public key, and the corresponding user credentials, are valid. Therefore, in practice we should avoid rules with monitoring periods larger than the group public key life. This especially relevant for rules with no time component (or infinite monitoring period), like the ones we present in the example~\ref{ntimesauth}. In general, whenever we allow key rotation there should always be some monitoring period in the rules, and this should be smaller than the group public key life.

\subsection{Possible de-anonymization attacks}

We discuss two possible de-anonymization attacks by the Issuer with the same vector: artificially reduce the number of members in a group. In the extreme case, a group could have a single user, then the collector could safely assume that all data receive comes from the same user.

Note that these attacks assume collusion between the collector receiving the data and the Issuer; it is not necessary that both entities are controlled by the same organization, they could be decoupled for extra safety. However, in practical terms it is unlikely that they are provably independent.


\subsubsection*{Ephemeral group public keys} \label{changinggroupkeys}

In our design we assume that group public keys can be rotated periodically by the Issuer, therefore there must be a way for a user to dynamically query the Issuer for the current valid group key, and possibly some of the next ones.

A possible attack that could be done by a malicious Issuer would consist in showing different group public keys to different users. In other words: trying to create many smaller groups to make de-anonymization easier.

Since the retrieval of the group public keys is anonymous there is no way for an Issuer to target specific users. Therefore, the only possibility would be to change the announced group public keys randomly, hoping that signers would not notice. Fortunately, this is easy to detect by a user: we can retrieve group public keys from the Issuer periodically and make sure they were not changed unexpectedly (before the announced expiration time). As an additional guarantee, in such case an attack was detected a signer could \textit{punish} the collector service in some way, for example by stopping all data collection.

\subsubsection*{Denying user group join}

Another possible de-anonymization attack would consist in only allowing specific users to join a group. 
Because the join operation is not anonymous, it requires user identifying information (e.g. a long-lived public key), this attack can be used to target a specific user or subset of users.

The attack, however, comes with the cost of stopping all data collection from users that were denied to join the group. In the case of targeting a single user, the collector would stop receiving data from everybody except the targeted user. Let us emphasize that the Issuer can deny a group join, but the user can verify whether he joined or not (see Sect.~\ref{annex:protocol:user}). In the case of join denial users are expected to stop sending data.

This attack cannot be prevented by design. However, the attack is extremely costly, tracking a user would imply losing all data from the rest of the users. Additionally, if sustained over a long period of time the attack might be noticed by users as some might publicly report the persisting failure to join.


\subsection{Unlinkability guarantees}

The standard notion of unlinkability in the Direct Anonymous Attestation setting requires that given a signature the probability that it belongs to a particular user must be equal for all the users.

With the proposed usage of the DAA basenames it is clear that our scheme does not fulfill this definition. First, if we find two unlinkable messages signed under the same basename we know for sure they were signed by different users. Second, users keep some local state (the set of already used basenames) which might be partially leaked to the collector. For example, if a user can send 100 messages (by choosing an unused nonce from 0 to 99), after enough nonces have been spent the possibilities for the remaining choices are reduced and the sequence of nonces might be predictable.

Still, if the amount of users contributing is large (e.g. tens of thousands), it is unfeasible to link messages from honest users with the proposed scheme. We do not foresee how the small amount of information that the collector can learn could be exploited.


\input{examples.tex}

%% file: examples.tex
\subsection{Examples}
\label{sec:examples}

Let us present some practical cases to illustrate how our system, via concrete rate-limiting rules definitions, can be used to prevent an attack. Whenever we mention basenames and rate-limiting rules, we refer to the formalization presented in Sect.~\ref{messagespecific}.


\subsubsection{GPS Location}

Suppose a location heatmap service that requires users to send their GPS position once in a 5 minutes interval. Let us call this service \textit{heatmap-service-1}.

Because the collector wants to respect the privacy of the users contributing data, the message must not contain PII, $uids$ or any other element that would allow to build a session for a particular user. Messages could be sent through Tor to remove network-level identifiers. A privacy-preserving message could look like this\footnote{Note that sampling often with high resolution in area with few users would allow for probabilistic linkage of records. Avoiding implicit linkage by the message (record) content is not an easy task, as discussed at length in~\cite{humanweb}.}:

\begin{verbatim}
    m = {
        latitute: 48.85034,
        longitude: 2.294694,
        timestamp: "2018/02/12T12:23",
        service: "heatmap-service-1"
    }
\end{verbatim}

The collector should have means of knowing which user is sending the message $m$. Therefore, an attacker could create thousands of messages to disrupt the service. To deter such an attack we must decide which rate-limiting we want to apply. In this case, let us assume we want to limit the number of messages per user to 1 every 5 minutes interval.

Formalizing it into a set of rate-limiting rules, and assuming the time units are in minutes, would result into one rule

\[digest_1(m) = \verb heatmap-service-1 \]
\[FK_1(m) = 5 \]
\[FN_1(m) = 1 \]

which when applied to the given sample could give as a result the following basename:

\begin{equation*}
    bsn_1 = \langle \texttt{heatmap-service-1} , \texttt{2018/02/12T12:20}, 0 \rangle
\end{equation*}

where \texttt{heatmap-service-1} is the result of the digest of the message (always constant in this case) and \texttt{2018/02/12T12:20} is the time rounded to 5 minute resolution (formatted for readability). The nonce is set to 0 as it does not apply in this scenario since $N=1$, i.e. 1 message per time period.




Note that our system does not aim to prevent an attacker sending bogus latitude and longitude. The goal is to ensure that the attacker will be only able to inject one bogus message every 5 minutes interval at best.

\subsubsection{Surveys}

This examples illustrates how we can set the rules to allow users to send anonymous surveys while ensuring each user can only submit a valid answer once. Anonize~\cite{anonize15} was designed to support this kind of use-cases (among other more elaborate). The message to be sent by users could look like this,

\begin{verbatim}
    m = {
        survey_id: "34ef2a",
        survey_data: {
            ...
        },
        timestamp: "2018/02/12T12:23",
        service: "survey-service-1"
    }
\end{verbatim}

We want to enforce that each user can only send the survey once. The rate-limiting rules for this use-case have no temporal aspect at all. To achieve this within the general ruleset framework from \ref{messagespecific} we can set $FK$ to return a very large constant, so that $\left\lfloor \frac{time}{FK(m)} \right\rfloor$ is always 0 in practice. The concrete rules would be

\[digest_1(m) = \verb survey-service-1 | m.surveyid \]
\[FK_1(m) = 2^{50} \]
\[FN_1(m) = 1 \]

which applied to the sample message would result in

\begin{equation*}
    bsn_1 = \langle \texttt{survey-service-1 | 34ef2a} , 0, 0 \rangle
\end{equation*}

We have chosen this example to showcase the scenario where only one message (the survery response) is needed. However, as we discussed in \ref{keyrotation}, even if we want a rule which is \textit{unlimited} in time, there will always be an implicit temporal limitation given by the group and user key rotation, after which theoretically the users will be able send messages again, since they will start with a fresh quota. Therefore, we should make a survey only be valid during a time span less than the duration of the group public key for which the messages have to be signed. Otherwise, the group key could rotate in the middle of a survey and a user could send a survey response twice and still be considered \textit{honest} under our rule definition.

\subsubsection{Query Logs}

The last example showcases a more complex case composed by multiple rate-limiting rules. Let us suppose we want to collect query-log pairs from users in order to improve the ranking of a search engine. This particular use case is vital for the proper functioning of Cliqz's search engine. The messages sent by users could look like this:

\begin{verbatim}
    m = {
        query: "hotel paris",
        landing_url: "https://www.booking.com/
                        city/fr/paris.htm",
        timestamp: "2018/02/12T12:23",
    }
\end{verbatim}

In this case, we want to enforce two rules at the same time:
\begin{enumerate}
    \item A user can only contribute 5 queries per day.
    \item For a specific query $q$ a user can only send a message once per day.
\end{enumerate}

This can be translated into the following rate-limiting ruleset, again assuming the time units are in minutes:

\[digest_1(m) = \verb query-log-service-1 \]
\[FK_1(m) = 24 \cdot 60 \]
\[FN_1(m) = 5 \]

\[digest_2(m) = \texttt{ query-log-service-2} | normalize(m.query) \]
\[FK_2(m) = 24 \cdot 60 \]
\[FN_2(m) = 1 \]

Applying both rules defined to the sample message could lead to something like:

\begin{equation*}
    bsn_{1} = \langle \texttt{ql-service-1} , \texttt{2018/02/12}, 3 \rangle
\end{equation*}
\begin{equation*}
    bsn_{2} = \langle \texttt{ql-service-2 | hotel paris} , \texttt{2018/02/12}, 0 \rangle
\end{equation*}

For the first rule we have a fixed digest and a date, and the nonce can be chosen between 0 and 4. For the second rule, the digest depends on the normalized query in the message and in this case the nonce can only be 0. Note that the digests of the rules have different static prefixes to avoid possible collisions.

They user would sign the message twice, once for each basename, and send it to the collector. After sending the message (assuming it is the first in the day) the user would still be able to send more messages for that day, 4 more, but not for the same query.

In order for these rules to be completely effective, the normalization applied in the digest function of the second rule should make sure that an attacker cannot use variations of a query to obtain different basenames for what it is esentially the same query. For instance, queries like \texttt{hotels in paris, hotel on paris, HoteL IN PARIS, hotels\textvisiblespace \textvisiblespace \textvisiblespace \textvisiblespace in paris}, etc. should all get normalized to the same digest\footnote{We can apply known transformations on the query like down-casing, trim spaces, bag-of-words, stemming, etc.}. This kind of attacks are use-case specific, as a rule-of-thumb, a designer should make sure that digests do not contain data that has not been sanitized and normalized.

%% file: protocol.tex
\section{Protocol}
\label{sec:protocol}

In this section we specify the protocol taking into consideration the concepts presented in the previous section. More specifically, we construct the \textbf{basenames} for rate-limiting rules as specified in \ref{messagespecific}, select the \textbf{nonces} and store related needed information as discussed in \ref{nonce}, and include operations for dynamic group and user \textbf{key rotation} as seen in \ref{keyrotation}.

For our protocol we use DAA as a cryptographic primitive, but without a TPM for the \textbf{Join} operation. We base our implementation on the concrete scheme presented in~\cite{cryptoeprint:2015:1246}, which is at the same time based on~\cite{Camenisch2004}.

Whenever we mention the operations \textbf{Setup}, \textbf{Join}, \textbf{Sign}, \textbf{Verify} or \textbf{ExtractTag} we are referring to the concrete DAA scheme operations, described in Annex~\ref{annex:1}. However, any DAA implementation that follows the semantics described in Sect.~\ref{preliminaries} might be used while keeping the same security and anonymity guarantees.

Before describing the protocol in detail, it is important to clarify the entities or actors that we consider in our system and how do they map to the entities in a DAA scheme,

\begin{itemize}
    \item \textbf{Signer}, \textbf{User} or \textbf{Client}: in our system this is any entity that sends data to a \textbf{Collector}, and corresponds to the \textbf{Signer} in DAA.
    \item \textbf{Issuer}: this corresponds to the Issuer in DAA.
    \item \textbf{Verifier} or \textbf{Collector}: in our system this is the entity that must verify the messages and decide whether they are valid or not.
\end{itemize}


\subsection*{RefreshGroupPublicKeys}

This operation is executed periodically by the user, to make sure the group public keys are still valid. The user should not send any message if the group public keys have expired.

\begin{itemize}
\item User anonymously requests to Issuer the list of group public keys.
\item Issuer returns a list of group public keys and time-to-live pairs $(gpk_0, expiry_0)$, $(gpk_1, expiry_1)$, $\ldots$, $(gpk_n, expiry_n)$, where $gpk_0$ and $expiry_0$ are the currently valid group public key and its expiration date, respectively, and the rest the $n$ next group public keys, ordered by increasing expiration.
\item User checks whether the list is consistent with previously stored information (Issuer has not changed any key before the announced expiration). If there was some unexpected change, \textit{punish} the collection process as defined in the concrete system and \textbf{abort}. (See attack in \ref{changinggroupkeys}).
\item User stores the retrieved list.
\item User executes \textbf{ObtainCredentials} on all the group public keys for which still does not have credentials.
\end{itemize}

\subsection*{ObtainCredentials}
\begin{itemize}
\item User executes the \textbf{Join} protocol with the issuer and a specific group public key $gpk$ to obtain credentials valid for that public key: $(usk, ucred) \leftarrow \boldsymbol{Join}(gpk)$
\item User stores the credentials $CREDS_u[gpk] \leftarrow (usk, ucred)$
\end{itemize}

\subsection*{RotateUserKeys}

This operation is executed periodically by the user.

\begin{itemize}
\item If $expiry_0 < \verb current_time $ then \textbf{abort}.
\item User replaces its current credentials for the ones of the new valid group public key $(usk, ucred) \leftarrow CREDS_u[gpk_1]$.
\item User invalidates the stored tags: $TAGS_u \leftarrow \emptyset$
\end{itemize}

\subsection*{RotateIssuerKeys}

This operation is executed periodically by the issuer.

\begin{itemize}
\item If $expiry_0 < \verb current_time $ then \textbf{abort}.
\item Issuer executes $(gsk, isk) \leftarrow \boldsymbol{Setup}()$, stores $(gsk, isk)$ and appends $(gpk, expiry)$ to the end of the public list.
\item Issuer rotates the public list $(gsk_i, expiry_i) \leftarrow (gsk_{i + 1}, expiry_{i + 1})$.
\item Issuer notifies the verifier of the new group key.
\item Verifier checks if the new $gpk$ is different from the previous one, and if so it invalidates its stored tags: $TAGS_v \leftarrow \emptyset$
\end{itemize}

\subsection*{SendMessage}

Input: a message $m$, user private key $usk$, user credentials $ucred$ and $n$ rate-limiting rules $(digest_1, FN_1, FK_1), ..., (digest_n, FN_n, FK_n)$.

\subsubsection*{User}
\begin{itemize}
\item User executes \textbf{RefreshGroupPublicKeys} if $expiration_0 \geq \verb current_time $.
\item User waits a reasonable randomized amount of time if some protocol operation was executed recently, including \textbf{SendMessage}, to avoid possible time correlations.
\item User computes $n$ pre-basenames on the message (all the components except the nonce). $prebsn_i \leftarrow \left < digest_i(m),  \left\lfloor \frac{time}{FK_i(m)} \right\rfloor \right >$.
\item User does a storage look-up for every $prebsn_i$: $(key_i, n_i) \leftarrow TAGS_u[prebsn_i]$. If it does not exist, initializes: $TAGS_u[prebsn_i] \leftarrow \big(key_i \xleftarrow{\$} \left\{0, 1 \right \}^\tau, 0\big)$. If some $n_i \geq FN_i(m)$ then \textbf{abort}.
\item User uses format-preserving encryption to create a fresh nonce for every pre-basename as $nonce_i \leftarrow FPE_{FN_i(m)}(key_i, n_i)$ and builds the basename as $bsn_i \leftarrow \left <digest_i(m), \left\lfloor \frac{time}{FK_i(m)} \right\rfloor, nonce_i \right >$

\item User increments the nonce for every stored pre-basename: $TAGS_u[prebsn_i] \leftarrow (key_i, n_i + 1)$

\item User produces $n$ signatures on $m$: $\sigma_i \leftarrow \boldsymbol{Sign}(usk, ucred, m, bsn_i)$

\item User sends $(m, \sigma_1, \ldots, \sigma_n, bsn_1, \ldots, bsn_n)$ to the verifier via an anonymous communication channel.

\end{itemize}

\subsubsection*{Verifier}
\begin{itemize}
\item Verifier checks that the all the basenames $bsn_i$ are correctly computed based on the message and the current time. If some is not $\textbf{abort}$.

\item Verifier checks validity of the signatures: if any $\boldsymbol{Verify}(gpk, m, bsn_i, \sigma_i)$ returns false then \textbf{abort}.

\item Verifier extracts the linkability tag for every signature $tag_i \leftarrow \boldsymbol{ExtractTag}(\sigma_i)$. If some $tag_i$ is present in $TAGS_v$ then \textbf{abort}.

\item Verifier inserts the all the tags in the storage: $TAGS_v \leftarrow TAGS_v \cup tag_i$.

\item Verifier \textbf{accepts} the message.
\end{itemize}

%% file: implementation.tex
\section{Implementation}

\subsection{Library}

For the implementation of our protocol we have used the \textit{Apache Milagro Crypto Library}~\cite{lib:milagro}. It is a self-contained, standalone library. We found the implemented operations fast enough for our purposes, so we did not look for other libraries. The library is implemented in C, and for our protocol we have chosen the same language.

For evaluating our system we have considered a server for the verifier/issuer part and a browser for the client, and ported the code to these platforms. For the server side we have built a Node.js~\cite{tilkov2010node} module via C bindings. For the browser, thanks to Emscripten~\cite{zakai2011emscripten} we have been able to compile the same protocol code to WebAssembly, achieving performance comparable to the native version. In Sect.~\ref{evaluation} we show some benchmark figures.

The code of the implemented Direct Anonymous Attestation primitives has been open-sourced~\cite{anonymouscreds}.

\subsection{Network anonymity}

In the introduction (Sect.~\ref{section:contribution}) we made the assumption that network anonymity is provided externally in some way, typically via a trusted VPN partner or some anonymity network like Tor.

For benchmarking our implementation we have chosen the latter. It provides better anonymity guarantees since nodes are under the control of many different organizations, so there is no single point of failure with regard to trust. Note that Tor, as any other system, is not free from de-anonymization attempts~\cite{kwon15, winter16}.

Using Tor, however, implies that users should be running the Tor client on their devices, which is not a very realistic assumption for general users. 
To overcome users not running a native Tor client we ported it~\cite{torjs} to WebAssembly, in a similar way as we did for the protocol implementation. This ensures that the Tor client logic (cryptographic operations, etc.) can be run in a browser. However, in some contexts in which this WebAssembly Tor version could be used (web pages, browser extensions) it is not possible to create raw TCP sockets as required by the native Tor client. In order to solve this issue  we have setup a Tor bridge and a WebSocket server adapter that proxies incoming WebSocket traffic to the internal Tor bridge.  On the browser side, we setup the Tor client to use our bridge, and use WebSockets to be able to communicate to the real Tor bridge. We believe this scenario should be equivalent to a native Tor client connecting to a Tor bridge via a WebSocket pluggable transport.

It is important to note that in the described setup the same organization would be in control of both the entry point (WebSocket Tor bridge) and the final destination of the data. This would allow for trivial correlation attacks, and result in no provable anonymization of the users. For that reason, we also tested the system with a WebSocket Tor Bridge ran by a third party. Coincidence or not, such bridge was already deployed as part of the \textit{Snowflake}~\cite{snowflake} pluggable transport project at Tor: all we needed was to replace our bridge address by the Snowflake one.

Let us remark that this approach is completely experimental and would need to be reviewed and audited before completely relying on its security and anonymity guarantees. Besides, it might be difficult in practice to find a suitable third-party WebSocket bridge. For example, it is unclear if it would be acceptable to use the mentioned Snowflake proxy, since it might put it under a load it was not prepared to handle.

Even with these concerns, we believe it might be an interesting option to explore in order to achieve network anonymity in a restricted environment like a browser extension or just regular web pages.

%% file: evaluation.tex
\section{Evaluation} \label{evaluation}

In this section we provide benchmarks for the protocol needed to guarantee protection against attackers (described in Sect.~\ref{sec:protocol}). We would like to stress that system presented in this paper is in production, supporting the anonymous data collection effort of few million users.

The experimental setup is as follows:
\begin{itemize}
    \item Issuer publishes group keys every 3 days.
    \item Clients fetch user credentials once every 3 days via Join protocol.
    \item Sign the messages using credentials fetched in the earlier step.
    \item Use Tor as the network layer for fetching the credentials and sending signed messages.
    \item Fixed payload sizes for request / response as shown in Table~\ref{table:3}.
\end{itemize}

Table~\ref{table:1} and~\ref{table:2} summarize the time overhead for client (users) and server (collector) respectively. The time is the average in milliseconds for different protocol operations. Network overhead is not considered.

For the client, the most expensive operation is Join which takes about 20ms, but it only runs once every 3 days.

\begin{table}[h!]
\caption{Average time spent on client-side operations. Also compares a native implementation of the client. Network overhead is not taken into account.}
\begin{tabular}{p{3cm}|c|c}
 Operation & Webassembly (ms) & Native (ms)\\ [0.5ex]
 \hline
  Join Group & 20 & 8.5 \\
  Sign message & 5 & 0.4 \\
\end{tabular}
\label{table:1}
\end{table}

For the server-side, the most expensive operation is Verify, which takes about 6.8ms per message per core on nominal hardware. This sets the throughput of the collector to approximately 150 messages per second per core, which is more than enough for our workload, we collect an average of 3 to 4 messages per minute per user. 

Note that verification does not need to be synchronous, a setup with queues, with high priority lanes if needed, will further help scalability.

\begin{table}[h!]
\caption{Average time spent on server-side operations. Network overhead is not taken into account.}
\begin{tabular}{p{5.7cm}|c}
 Operation & Time spent (ms)\\ [0.5ex]
 \hline
  Generate group keys  & 2.3\\
  Generate credentials & 1.6 \\
  Verify message & 6.8\\
\end{tabular}
\label{table:2}
\end{table}

The implementation on both server and client takes care of padding the messages to fixed sizes in order to prevent fingerprinting by measuring payload sizes. Table~\ref{table:3} shows the size of the payload at each step of request-response.

\begin{table}[h!]
\caption{Payload sizes}
\begin{tabular}{p{2.5cm}|c|c}
 Operation & Request size (KB) & Response size (KB)\\ [0.5ex]
 \hline
  Fetch group keys & -- &  5.04 \\
  Fetch credentials & 0.95 & 0.477 \\
  Signed Message & 16.384 & 0.021 \\
\end{tabular}
\label{table:3}
\end{table}

Based on these payload sizes, Table~\ref{table:4} compares the $95^{th}$ percentile network latency when communicating with end-points without Tor, with end-points over Tor, end-points as onion services.

\begin{table}[h!]
\caption{Comparison of $95^{th}$ percentile network latency for Send message operation using Tor network.}
\begin{tabular}{p{5.5cm}|c }
 Setup & Latency (seconds) \\
 \hline
  Without Tor & 0.229 \\
  With Tor and endpoints as normal services & 3.25 \\
  With Tor and endpoints as onion services & 2.57 \\
\end{tabular}
\label{table:4}
\end{table}

Although the evaluation shows that latency increases ~10x when communication is through Tor, is a very good trade-off taking into consideration the benefits it provides. Also, latency overhead is only important for synchronous communication, which is rarely the case for data collection.

%% file: conclusion.tex
\section{Conclusions}

We have presented a system to effectively and efficiently prevent attackers from polluting data collection by abusing the anonymity that guarantees the users' privacy.

Thanks to the presented system a collector can define a normative space, which is enforced cryptographically using Direct Anonymous Attestation (DAA): any message that infringes any of the defined rate-limiting rules can be detected, and consequently, dropped. The anonymity of users sending data is preserved at all times.

The normative space defined by the collector has a set of rate-limiting rules on multiple dimensions: time, content and multipliers, so that a wide-range of use-cases can be accommodated. For instance, a collector could require that record of type $x$ can be sent once per hour,  and that records of type $y$ can be sent 5 times a day if an only if its content matches a certain pattern. The defined rate-limiting rules can be formally mapped to a basename on DAA, which will offer the required cryptographic guarantees.

We also present a description of all components needed to implement and deploy the system, including an evaluation of its performance.

We hope that this work will help those organizations hesitant to use anonymous data collection becasue of potential data pollution. We demonstrate that is possible defend against adversarial attacks while maintaining anonymity, and consequently, privacy. Protection against sybil attacks is limited to the difficulty of creating/renewing identities, which is domain specific. Protection against single identities, however, is entirely covered.

We believe that our contribution demonstrates the feasibility of deploying anonymous data collection systems. Once the fear of data pollution is out of the picture, there is no reason to not collect the data anonymously, unless, of course, convenience is more valued than the users' privacy.




%% file: annex_1.tex
\section{Concrete DAA scheme} \label{annex:1}

\subsection{Preliminaries}

\subsubsection*{Bilinear Maps}
Let $\mathbb{G}_1$, $\mathbb{G}_2$ and $\mathbb{G}_T$ be a bilinear group of prime order $q$ with generators $g_1, g_2$, $e : \mathbb{G}_1 \times \mathbb{G}_2 \rightarrow \mathbb{G}_T $ a bilinear map and $H_1 : \{0,1\}^* \rightarrow \mathbb{G}_1$, $H : \{0, 1\}^* \rightarrow \mathbb{Z}_q$ cryptographic hash functions.

\subsubsection*{Signatures of Knowledge}

A zero-knowledge proof is a method by which one party (prover) can convince another party (verifier) that a statement is true, without disclosing any information apart from that fact.

We will use the definitions and notation introduced by Camenisch and Stadler \cite{Camenisch1997} to present the concept of \textit{Signatures of Knowledge}, which are based on zero-knowledge proofs.

Let us introduce the notation
\[ SPK\{(x) : y = g^x\}(m)\]
denoting a signature of knowledge of the discrete logarithm of $y$ on the message $m$. This signature can be computed if the secret key $x = log_g(y)$ is known, by choosing $r$ at random from $\mathbb{Z}_q$ and computing $c$ and $s$ according to

\[c \leftarrow H(m || y || g || g^r) \qquad\text{and}\qquad s \leftarrow r - cx \mod q\]

A valid signature will satisfy

\[c = H(m || y || g || g^sy^c)\]

Note that $x$ is not revealed, the proof itself just consists of the pair $(c, s)$. The elements $y$, $g$ and $m$ are considered public.

A variant which we will also employ is the following, proving the equality of the logarithm of two elements:

\[ SPK\{(x) : y = a^x \wedge z = b^x\}(m)\]

where $a$ and $b$ are generators of $G$. Again this can be computed with knowledge of $x$, by choosing $r$ at random from $\mathbb{Z}_q$ and computing $c$ and $s$ according to

\[c \leftarrow H(m || y || z || a || b || a^r || b^r) \qquad\text{and}\qquad s \leftarrow r - cx \mod q\]

Here a valid signature will satisfy

\[c = H(m || y || z || a || b || a^sy^c || b^sz^c)\]

These two types of signatures of knowledge (or zero-knowledge proofs) will suffice for the protocol.

\subsection{Scheme}

We use the scheme described in \cite{cryptoeprint:2015:1246}, which is at the same time based on \cite{Camenisch2004}.

\subsection*{Setup}
Used by the issuer to create a key pair of the CL-signature scheme.
\begin{itemize}
\item Choose $x, y \xleftarrow{\$} \mathbb{Z}_q$, and set $X \leftarrow g^x_2$, $Y \leftarrow g^y_2$.
\item Compute the proof: $\pi \xleftarrow{\$} SPK \left\{(x, y) : X = g^x_2 \wedge Y = g^y_2 \right\}$.
\item Output a key pair as $\left < isk = (x, y), gpk = (X, Y, \pi) \right >$ where $isk$ is the issuer secret key and $gpk$ the corresponding group public key.
\end{itemize}

\subsection*{Join}

User authenticates with her long lived PKI ($u_{pk}, u_{sk}$) and, if the issuer allows, obtains a credential that subsequently enables the user to create signatures.

\subsubsection*{User}
\begin{itemize}
\item Choose $gsk \xleftarrow{\$} \mathbb{Z}_q$.
\item Compute $c \leftarrow H(X, Y, \pi, u_{pk})$
\item Set $Q = g^{gsk}_1$ and compute $\pi_1 \xleftarrow{\$} SPK \left\{(gsk) : Q = g^{gsk}_1 \right\}(c)$
\item Store $gsk$ and send $msg_{join} \leftarrow (u_{pk}, Q, \pi_1)$ and $sig_{msg} \leftarrow sign(msg_{join}, u_{sk})$ to the issuing service.
\end{itemize}

\subsubsection*{Issuer}
\begin{itemize}
\item Set $(x, y) \leftarrow isk$ and $(X, Y) \leftarrow gpk$.
\item Verify $\pi_1$, the signature $sig_{msg}$ and check whether user $u_{pk}$ is registered.
\item If there are stored (previously generated) credentials for $u_{pk}$, and user is not allowed to get new ones, send those to the user. Otherwise:
\item Choose $r \xleftarrow{\$} \mathbb{Z}_q$ and compute $a \leftarrow g^r_1$, $b \leftarrow a^y$, $c \leftarrow a^x \cdot Q^{r x y}$, $d \leftarrow Q^{r y}$.
\item Compute $\pi_2 \xleftarrow{\$} SPK \left\{(t) : b = g^{t}_1 \wedge d = Q^t \right\}$
\item Store the credentials for user $u_{pk}$ as $(a, b, c, d, \pi_2)$.
\item Transmit credentials to user.
\end{itemize}

\subsubsection*{User}
\label{annex:protocol:user}

\begin{itemize}
\item Verify $\pi_2$.
\item Verify the credentials as $a \neq 1$, $e(a, Y) = e(b, g_2)$, and $e(c, g_2) = e(a \cdot d, X)$.
\item Complete the join by appending $(a, b, c, d)$ to the already stored $gsk$.
\end{itemize}

\subsection*{Sign}
Input: user secret key $gsk$, user credentials $(a, b, c, d)$, a basename $bsn$, and a message $m$.

\begin{itemize}
\item Choose $r \xleftarrow{\$} \mathbb{Z}_q$ and set $(a', b', c', d') \leftarrow (a^r, b^r, c^r, d^r)$.
\item Compute $\pi \xleftarrow{\$} SPK \{(gsk) : tag = H_1(bsn)^{gsk} \wedge  d' = b'^{gsk}\}(m, bsn)$.
\item The signature is $\sigma \leftarrow (a', b', c', d', \pi, tag)$.

\end{itemize}

\subsection*{Verify}
Input: user secret key $gpk$, a basename $bsn$, a message $m$ and a candidate signature $\sigma$.

\begin{itemize}
\item Parse $\sigma$ as $(a, b, c, d, \pi, tag)$.
\item Verify $\pi$ with respect to $(m, bsn)$.
\item Check $a \neq 1$, $e(a, Y) = e(b, g_2)$, and $e(c, g_2) = e(a \cdot d, X)$.
\item Accept signature as valid.
\end{itemize}

\subsection*{ExtractTag}
Input: a valid signature $\sigma$.
\begin{itemize}
    \item Parse $\sigma$ as $(a, b, c, d, \pi, tag)$
    \item Return $tag$
\end{itemize}